\documentclass[twocolumn,showpacs,preprintnumbers,amsmath,amssymb,superscriptaddress,floatfix]{revtex4}

\usepackage{graphicx}  
\usepackage{dcolumn}   
\usepackage{bm}        

\newcommand{\met}{\,/\!\!\!\!E_{T}}

\begin{document}



\title{Search for Higgs bosons predicted in two-Higgs-doublet models
       via decays to tau lepton pairs in 1.96 TeV $p\bar{p}$ collisions}

\date{\today}

\begin{abstract}
We present the results of a search for Higgs bosons predicted in
two-Higgs-doublet models, in the case where the Higgs bosons decay to
tau lepton pairs, using 1.8 fb$^{-1}$ of integrated luminosity of
$p\bar{p}$ collisions recorded by the CDF II experiment at the
Fermilab Tevatron.  Studying the mass distribution in events
where one or both tau leptons decay leptonically, no evidence for a
Higgs boson signal is observed.  The result is used to infer exclusion
limits in the two-dimensional space of $\tan\beta$ versus
$m_A$ (the ratio of the vaccum expectation values of the two Higgs doublets
and the mass of the pseudoscalar boson, respectively).
\end{abstract}


\affiliation{Institute of Physics, Academia Sinica, Taipei, Taiwan 11529, Republic of China} 
\affiliation{Argonne National Laboratory, Argonne, Illinois 60439} 
\affiliation{University of Athens, 157 71 Athens, Greece} 
\affiliation{Institut de Fisica d'Altes Energies, Universitat Autonoma de Barcelona, E-08193, Bellaterra (Barcelona), Spain} 
\affiliation{Baylor University, Waco, Texas  76798} 
\affiliation{Istituto Nazionale di Fisica Nucleare Bologna, $^y$University of Bologna, I-40127 Bologna, Italy} 
\affiliation{Brandeis University, Waltham, Massachusetts 02254} 
\affiliation{University of California, Davis, Davis, California  95616} 
\affiliation{University of California, Los Angeles, Los Angeles, California  90024} 
\affiliation{University of California, San Diego, La Jolla, California  92093} 
\affiliation{University of California, Santa Barbara, Santa Barbara, California 93106} 
\affiliation{Instituto de Fisica de Cantabria, CSIC-University of Cantabria, 39005 Santander, Spain} 
\affiliation{Carnegie Mellon University, Pittsburgh, PA  15213} 
\affiliation{Enrico Fermi Institute, University of Chicago, Chicago, Illinois 60637}
\affiliation{Comenius University, 842 48 Bratislava, Slovakia; Institute of Experimental Physics, 040 01 Kosice, Slovakia} 
\affiliation{Joint Institute for Nuclear Research, RU-141980 Dubna, Russia} 
\affiliation{Duke University, Durham, North Carolina  27708} 
\affiliation{Fermi National Accelerator Laboratory, Batavia, Illinois 60510} 
\affiliation{University of Florida, Gainesville, Florida  32611} 
\affiliation{Laboratori Nazionali di Frascati, Istituto Nazionale di Fisica Nucleare, I-00044 Frascati, Italy} 
\affiliation{University of Geneva, CH-1211 Geneva 4, Switzerland} 
\affiliation{Glasgow University, Glasgow G12 8QQ, United Kingdom} 
\affiliation{Harvard University, Cambridge, Massachusetts 02138} 
\affiliation{Division of High Energy Physics, Department of Physics, University of Helsinki and Helsinki Institute of Physics, FIN-00014, Helsinki, Finland} 
\affiliation{University of Illinois, Urbana, Illinois 61801} 
\affiliation{The Johns Hopkins University, Baltimore, Maryland 21218} 
\affiliation{Institut f\"{u}r Experimentelle Kernphysik, Universit\"{a}t Karlsruhe, 76128 Karlsruhe, Germany} 
\affiliation{Center for High Energy Physics: Kyungpook National University, Daegu 702-701, Korea; Seoul National University, Seoul 151-742, Korea; Sungkyunkwan University, Suwon 440-746, Korea; Korea Institute of Science and Technology Information, Daejeon, 305-806, Korea; Chonnam National University, Gwangju,500-757, Korea; Chonbuk National University, Jeonju 561-756, Korea} 
\affiliation{Ernest Orlando Lawrence Berkeley National Laboratory, Berkeley, California 94720} 
\affiliation{University of Liverpool, Liverpool L69 7ZE, United Kingdom} 
\affiliation{University College London, London WC1E 6BT, United Kingdom} 
\affiliation{Centro de Investigaciones Energeticas Medioambientales y Tecnologicas, E-28040 Madrid, Spain} 
\affiliation{Massachusetts Institute of Technology, Cambridge, Massachusetts  02139} 
\affiliation{Institute of Particle Physics: McGill University, Montr\'{e}al, Qu\'{e}bec, Canada H3A~2T8; Simon Fraser University, Burnaby, British Columbia, Canada V5A~1S6; University of Toronto, Toronto, Ontario, Canada M5S~1A7; and TRIUMF, Vancouver, British Columbia, Canada V6T~2A3} 
\affiliation{University of Michigan, Ann Arbor, Michigan 48109} 
\affiliation{Michigan State University, East Lansing, Michigan  48824}
\affiliation{Institution for Theoretical and Experimental Physics, ITEP, Moscow 117259, Russia} 
\affiliation{University of New Mexico, Albuquerque, New Mexico 87131} 
\affiliation{Northwestern University, Evanston, Illinois  60208} 
\affiliation{The Ohio State University, Columbus, Ohio  43210} 
\affiliation{Okayama University, Okayama 700-8530, Japan} 
\affiliation{Osaka City University, Osaka 588, Japan} 
\affiliation{University of Oxford, Oxford OX1 3RH, United Kingdom} 
\affiliation{Istituto Nazionale di Fisica Nucleare, Sezione di Padova-Trento, $^z$University of Padova, I-35131 Padova, Italy} 
\affiliation{LPNHE, Universite Pierre et Marie Curie/IN2P3-CNRS, UMR7585, Paris, F-75252 France} 
\affiliation{University of Pennsylvania, Philadelphia, Pennsylvania 19104}
\affiliation{Istituto Nazionale di Fisica Nucleare Pisa, $^{aa}$University of Pisa, $^{bb}$University of Siena and $^{cc}$Scuola Normale Superiore, I-56127 Pisa, Italy} 
\affiliation{University of Pittsburgh, Pittsburgh, Pennsylvania 15260} 
\affiliation{Purdue University, West Lafayette, Indiana 47907} 
\affiliation{University of Rochester, Rochester, New York 14627} 
\affiliation{The Rockefeller University, New York, New York 10021} 
\affiliation{Istituto Nazionale di Fisica Nucleare, Sezione di Roma 1, $^{dd}$Sapienza Universit\`{a} di Roma, I-00185 Roma, Italy} 

\affiliation{Rutgers University, Piscataway, New Jersey 08855} 
\affiliation{Texas A\&M University, College Station, Texas 77843} 
\affiliation{Istituto Nazionale di Fisica Nucleare Trieste/Udine, I-34100 Trieste, $^{ee}$University of Trieste/Udine, I-33100 Udine, Italy} 
\affiliation{University of Tsukuba, Tsukuba, Ibaraki 305, Japan} 
\affiliation{Tufts University, Medford, Massachusetts 02155} 
\affiliation{Waseda University, Tokyo 169, Japan} 
\affiliation{Wayne State University, Detroit, Michigan  48201} 
\affiliation{University of Wisconsin, Madison, Wisconsin 53706} 
\affiliation{Yale University, New Haven, Connecticut 06520} 
\author{T.~Aaltonen}
\affiliation{Division of High Energy Physics, Department of Physics, University of Helsinki and Helsinki Institute of Physics, FIN-00014, Helsinki, Finland}
\author{J.~Adelman}
\affiliation{Enrico Fermi Institute, University of Chicago, Chicago, Illinois 60637}
\author{T.~Akimoto}
\affiliation{University of Tsukuba, Tsukuba, Ibaraki 305, Japan}
\author{B.~\'{A}lvarez~Gonz\'{a}lez$^t$}
\affiliation{Instituto de Fisica de Cantabria, CSIC-University of Cantabria, 39005 Santander, Spain}
\author{S.~Amerio$^z$}
\affiliation{Istituto Nazionale di Fisica Nucleare, Sezione di Padova-Trento, $^z$University of Padova, I-35131 Padova, Italy} 

\author{D.~Amidei}
\affiliation{University of Michigan, Ann Arbor, Michigan 48109}
\author{A.~Anastassov}
\affiliation{Northwestern University, Evanston, Illinois  60208}
\author{A.~Annovi}
\affiliation{Laboratori Nazionali di Frascati, Istituto Nazionale di Fisica Nucleare, I-00044 Frascati, Italy}
\author{J.~Antos}
\affiliation{Comenius University, 842 48 Bratislava, Slovakia; Institute of Experimental Physics, 040 01 Kosice, Slovakia}
\author{G.~Apollinari}
\affiliation{Fermi National Accelerator Laboratory, Batavia, Illinois 60510}
% ...

\collaboration{CDF Collaboration\footnote{With visitors from $^a$University of Massachusetts Amherst, Amherst, Massachusetts 01003,
$^b$Universiteit Antwerpen, B-2610 Antwerp, Belgium, 
$^c$University of Bristol, Bristol BS8 1TL, United Kingdom,
$^d$Chinese Academy of Sciences, Beijing 100864, China, 
$^e$Istituto Nazionale di Fisica Nucleare, Sezione di Cagliari, 09042 Monserrato (Cagliari), Italy,
$^f$University of California Irvine, Irvine, CA  92697, 
$^g$University of California Santa Cruz, Santa Cruz, CA  95064, 
$^h$Cornell University, Ithaca, NY  14853, 
$^i$University of Cyprus, Nicosia CY-1678, Cyprus, 
$^j$University College Dublin, Dublin 4, Ireland,
$^k$University of Edinburgh, Edinburgh EH9 3JZ, United Kingdom, 
$^l$University of Fukui, Fukui City, Fukui Prefecture, Japan 910-0017
$^m$Kinki University, Higashi-Osaka City, Japan 577-8502
$^n$Universidad Iberoamericana, Mexico D.F., Mexico,
$^o$University of Iowa, Iowa City, IA  52242,
$^p$Queen Mary, University of London, London, E1 4NS, England,
$^q$University of Manchester, Manchester M13 9PL, England, 
$^r$Nagasaki Institute of Applied Science, Nagasaki, Japan, 
$^s$University of Notre Dame, Notre Dame, IN 46556,
$^t$University de Oviedo, E-33007 Oviedo, Spain, 
$^u$Texas Tech University, Lubbock, TX  79609, 
$^v$IFIC(CSIC-Universitat de Valencia), 46071 Valencia, Spain,
$^w$University of Virginia, Charlottesville, VA  22904,
$^x$Bergische Universit\"at Wuppertal, 42097 Wuppertal, Germany,
$^{ff}$On leave from J.~Stefan Institute, Ljubljana, Slovenia, 
}}
\noaffiliation

\pacs{12.60.Fr, 12.60.Jv, 13.85.Rm, 14.80.Cp} 



\maketitle


Understanding the origin of electroweak symmetry breaking is one of
the central goals of particle physics.  The Higgs
mechanism~\cite{higgs} in the standard model (SM) provides a possible
explanation, but the calculated mass of the Higgs boson suffers from
large radiative corrections.  Remedies for this problem such as
supersymmetry~\cite{susy} require at least two Higgs
doublets~\cite{2hdm} and result in a more complicated Higgs boson
sector than that of the SM.  The minimal supersymmetric standard model
(MSSM)~\cite{MSSM} predicts the existence of three neutral Higgs
bosons.  The MSSM is an example of a Type II two-Higgs-doublet model
(Type II 2HDM) in which there is a light scalar $h$, a heavy scalar
$H$, and a pseudoscalar $A$.  The masses of these states are governed
mainly by two parameters in the theory, usually taken to be
$\tan\beta$, the ratio of the vacuum expectation values of the two
Higgs doublets, and $m_A$, the mass of the pseudoscalar.

In $p\bar{p}$ collisions at 1.96 TeV center of mass energy at the
Fermilab Tevatron, Type II 2HDM Higgs bosons would be predominantly
produced by gluon-gluon fusion through a $b$ quark
loop~\cite{ggfusion} or by $b\bar{b}$ fusion~\cite{bbfusion}.  The
couplings and masses of the Higgs bosons are such that if $\tan\beta$
is greater than about 20 and $m_A$ is smaller (greater) than about 125
GeV/$c^2$, one finds that the $h$ ($H$) and $A$ are degenerate in mass
to within a few GeV/$c^2$, and are produced with a cross section
proportional to $\tan^2\beta$, while the production of $H$ ($h$) is
suppressed.  

These $\tan^2\beta$-enhanced production cross sections can be in the
range 0.1 -- 10 pb depending on the Higgs boson masses, and are orders
of magnitude greater than the corresponding ones for a SM Higgs and
also the more familiar associated production modes of a SM Higgs boson
with a vector boson.  The Higgs bosons decay to fermion pairs with a
partial width proportional to the fermion mass squared; thus the
decays $\phi\to b\bar{b}$ and $\phi\to\tau^+\tau^-$ (with
$\phi=h,A,H$) predominate, with the branching ratio to $b\bar{b}$
approximately 90\% and the branching ratio to $\tau^+\tau^-$ about 9\%
for $m_A >$ 100 GeV/$c^2$.

This Letter presents the results of a search for the production of
Higgs bosons in Type II 2HDM such as the MSSM,
using data collected with the CDF II detector at the Fermilab Tevatron
$p\bar{p}$ collider corresponding to 1.8~fb$^{-1}$ of integrated
luminosity.  Full details of the analysis are available
elsewhere~\cite{cris}; this result supersedes our previously published
result~\cite{Jan07}, and is similar to the search performed by the
D0 Collaboration~\cite{D0}.  The analysis is sensitive to a region of 
MSSM parameter space which is complementary to that explored by the 
LEP 2 experiments.~\cite{LEP2}

The analysis presented here uses the tau pair decay modes, since it is
possible to efficiently trigger on and reconstruct the leptons in 
decays of the tau lepton to $e\nu\bar{\nu}$ or $\mu\nu\bar{\nu}$.
Indeed, despite the $10\times$ larger branching ratio to $b\bar{b}$, the
search in the tau mode is more sensitive because the SM background
is much smaller.


CDF II~\cite{cdf} is a general-purpose detector with an overall
cylindrical geometry surrounding the $p\bar{p}$ interaction region.
The three-dimensional trajectories of charged particles produced in
$p\bar{p}$ collisions are measured at small radii ($<$ 30 cm) using
multiple layers of silicon microstrip detectors, and at outer radii
($>$ 30 cm) with a multi-wire drift chamber.  The tracking system is
inside a solenoidal magnet with uniform 1.4 T magnetic field oriented
along the beam direction.  Outside the solenoid are the
electromagnetic and hadronic calorimeters, which are segmented in
pseudorapidity ($\eta$) and azimuth in a projective ``tower''
geometry~\cite{ref-eta}. A set of strip and wire chambers located at a
depth of six radiation lengths aids in identifying photons and
electrons from the electromagnetic shower shape.  Muons are identified
by a system of drift chambers and scintillators placed outside the
calorimeter steel, which acts as an absorber for hadrons.  The
integrated luminosity of the $p\bar{p}$ collisions is measured using
\v{C}erenkov luminosity counters~\cite{ref-lum}.


We seek events with tau pairs where one or both taus decay
leptonically (excluding $e^+e^-$ and $\mu^+\mu^-$ which suffer from
excessive background from $Z/\gamma^*$ production).  These final
states are denoted $e+\tau$, $\mu+\tau$, and $e+\mu$ (where ``$\tau$''
here means the reconstructed hadronic part of a tau decay).  Events
with a high-$p_T$ (8 GeV/$c$ or more) $e$ or $\mu$ candidate plus a
high-$p_T$ charged track (5 GeV/$c$ or more) or a second $e$ or $\mu$
(4 GeV/$c$ or more) are identified using high-speed trigger
electronics and are recorded for later analysis.  The performance of
the trigger and lepton identification algorithms is described in
detail elsewhere~\cite{LT,lepton-ID}.


The reconstruction of hadronic decays of tau leptons~\cite{tau-ID} relies
on defining tau signal and isolation region cones centered around seed
tracks having $p_T > $ 6 GeV/$c$; we demand one or three charged
tracks in the tau cone, and include in addition any $\pi^0$
candidates.  The main way to discriminate between hadronic tau lepton
decays and hadronic jets from quantum chromodynamic processes is to
demand no additional charged tracks or $\pi^0$ candidates in an
isolation annulus outside the tau signal cone but within 30$^\circ$ of
the tau seed track.  The half-angle of the tau signal cone decreases
with increasing visible tau energy due to the Lorentz boost of the tau
lepton, further aiding the discrimination of taus from jets.
Additional suppression of hadronic jets comes from imposing a mass
requirement on the tau candidate decay products.  Electrons and muons
are removed using information from the calorimeters and muon
detectors.

To select the $e+\tau$ and $\mu+\tau$ events we require an isolated
$e$ or $\mu$ with $p_T >$ 10 GeV/$c$, and a $\tau$ with visible hadronic
decay products with total $p_T >$ 15 GeV/$c$ (20 GeV/$c$ for
three-charged-pion decays).  For the $e+\mu$ channel we require one
lepton to have $p_T >$ 10 GeV/$c$ and the other to have $p_T >$ 6 GeV/$c$.  


The main SM contributions to the selected event sample include
$Z/\gamma^*\to\tau^+\tau^-$, and $W+$jet events where $W\to\ell\nu$
(with $\ell=e,\mu$) and the hadronic jet is misidentified as a
hadronically decaying $\tau$.  The $W+$jet events are largely removed
by requiring that the missing transverse energy $\vec{\met}$ not point along
the direction opposite the momentum of the $\ell+\tau$ system.  The
remaining $W$+jet background, and all other background stemming from
jets misreconstructed as taus (``fakes'') is estimated from events
recorded with a jet trigger.  There are small contributions from $Z\to
e^+e^-$, $Z\to\mu^+\mu^-$, diboson, and $t\bar{t}$ production.

The acceptances for signal and the non-fake backgrounds are estimated
from samples of simulated events produced by the \textsc{pythia} event
generator~\cite{pythia} with \textsc{cteq5l}~\cite{cteq5l} parton
distribution functions.  The Higgs boson widths and masses are those
for an MSSM model with $\tan\beta$ = 50.  Tau decays are simulated by
the \textsc{tauola} package~\cite{tauola}. A
\textsc{geant}-based~\cite{geant} model simulates the interactions of
all final state particles in the detector.

Table \ref{tab-bkg} shows the mean expected contributions of SM sources,
and the number of observed events in the three channels.  The uncertainties
listed include all systematic effects discussed below, including correlations.

\begin{table}[t!]
  \caption{Mean expected SM backgrounds and observed numbers of selected events in the
           final sample. The uncertainties include all systematic effects, some of which
           are correlated.}
  \begin{tabular}{lccc}\hline\hline
     Background                  &        $e+\mu$        & $e+\tau$      &   $\mu+\tau$            \\ \hline
     $Z/\gamma^*\to\tau^+\tau^-$ &     605 $\pm$  51   &   1378 $\pm$ 117  &  1353 $\pm$  116 \\
     $Z/\gamma^*\to e^+ e^-$     &     1.5 $\pm$  1.2  &     70 $\pm$ 10   &     negl.        \\
     $Z/\gamma^*\to \mu^+\mu^-$  &    17.9 $\pm$  4.5  &      negl.        &   107 $\pm$  13  \\
     dibosons                    &    11.4 $\pm$  3.7  &    4.2 $\pm$ 2.1  &   3.3 $\pm$  1.8 \\
     $t\bar{t}$                  &     9.1 $\pm$  3.3  &    4.0 $\pm$ 2.1  &   3.3 $\pm$  1.9 \\
     $W$+jet, multijet           &    57.1 $\pm$  13.5 &    467 $\pm$ 73   &   285 $\pm$  46  \\
     Total                       &     702 $\pm$  55   &   1922 $\pm$ 141  &  1752 $\pm$  129 \\ \hline
     Observed                    &      726            &   1979            &     1666         \\ \hline \hline
  \end{tabular}
  \label{tab-bkg}
\end{table}

\begin{figure}
  \begin{center}
     \includegraphics[width=\columnwidth]{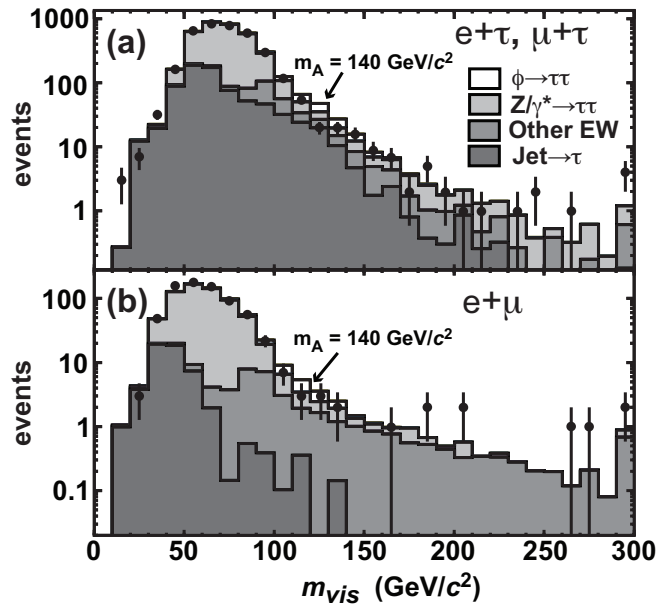} \\
  \end{center}
  \caption{Observed and predicted distributions of $m_{vis}$ for the $e+\tau$ and 
           $\mu+\tau$ channels (a) and $e+\mu$ channel (b).  The predicted 
           signal distribution (for $\phi=h/A/H$) corresponds to that for $m_A =$ 140 GeV/$c^2$ signal 
           assuming a value of the cross section excluded at 95\% C.L.  Note that in 
           each plot the last bin is an overflow bin.  Here ``Other EW'' refers to 
           all SM backgrounds other than $Z/\gamma^*\to\tau\tau$ and background arising
           from jet$\to\tau$ misidentification.}
  \label{fig-Mvis}
\end{figure}

To discriminate a Higgs boson signal from the backgrounds, we perform
a binned likelihood fit of background and signal to the observed
distribution of the ``visible mass'' $m_{vis}$, derived from the sum
of the observed lepton four-momenta and the missing transverse energy.
The observed distribution of this quantity is dominated by the effects
of the missing neutrino energy in the tau decays and experimental
resolution.  Figure~\ref{fig-Mvis} shows the observed and fit
distributions for the search channels, including the contribution from
a Higgs boson signal as described below.

Various uncertainties limit the sensitivity of our search.  The one
with the largest effect is due to the imprecisely known tau energy.
The distribution of the observed transverse momentum of the $\tau$ in
$W\to\tau\nu$ events constrains the ratio of the reconstructed tau
energy in the observed events to that in the simulation to less than
1\%, but the residual uncertainty allows for shifts in the background
$m_{vis}$ spectra mimicking a Higgs boson signal, particularly for the
lower masses considered ($m_A <$ 140 GeV/$c^2$).  At larger Higgs
boson masses the search sensitivity is limited more by other
systematic effects considered, including the lepton trigger and
identification uncertainties (2.4\% for electrons, 2.7\% for muons,
and 4.2\% for hadronically decaying taus), the uncertainty in the
integrated luminosity (6\%), $Z\to\tau^+\tau^-$ cross section (2.2\%),
and Higgs boson production cross section (5.7\%)~\cite{CTEQ6}.

We represent all the systematic uncertainties by Gaussian-constrained
nuisance parameters in the likelihood, and eliminate these parameters
by maximizing the likelihood with respect to them.  This procedure is
numerically nearly identical to eliminating them by Bayesian
marginalization with a Gaussian prior density, which takes much longer
to compute.

The resulting likelihood is calculated as a function of the Higgs
boson signal cross section times branching ratio to tau pairs
$\sigma\cdot B$, and then converted to a posterior density in
$\sigma\cdot B$ assuming a uniform prior density.  We exclude with
95\% C.L. any $\sigma\cdot B$ above which 5\% of the posterior
probability lies.


The likelihood as a function of $\sigma\cdot B$ reveals no evidence
for the presence of a Higgs boson signal, and all nuisance parameters
remain consistent with their nominal values.  Figure~\ref{fig-Mvis} and
other kinematic distributions not shown in this Letter all reveal
excellent agreement of the observed distributions with the predictions.
We therefore proceed to use the null result to infer upper limits on the
Higgs boson production cross section.

Table~\ref{tab-limits} lists the observed 95\% C.L. upper limits on
$\sigma\cdot B$, and the median upper limits expected under the null
hypothesis. Figure~2 depicts the results, including $\pm 1\sigma$ and
$\pm 2\sigma$ ranges for the expected limits. We note that these
bounds would apply to any scalar with similar production kinematics
decaying to tau pairs.  The Higgs boson signal distribution shown in
Fig.~\ref{fig-Mvis} corresponds to that excluded at 95\% C.L. for $m_A
= 140$ GeV/$c^2$.

We can interpret the upper limits on $\sigma\cdot B$ in the context of
the MSSM parameters $\tan\beta$ and $m_A$.  The resulting excluded
regions are shown in Fig.~\ref{fig-tanb-ma} for various assumptions
about the sign of the higgsino mass parameter $\mu$ and two extremes
for the nature of scalar top mixing~\cite{carena}, denoted $m_h^{max}$
and ``no mixing.''  The excluded regions are the most stringent
published to date in the high $\tan\beta$ region, and are remarkably
insensitive to changes in theoretical assumptions due to cancellation
of effects in the Higgs boson production and decay~\cite{carena2}.


In summary, we have used a sample of data from the Tevatron collider
recorded by the CDF II detector corresponding to 1.8 fb$^{-1}$ of
integrated luminosity to search for Higgs bosons predicted in
two-Higgs-doublet models, via the Higgs boson decays to tau lepton
pairs.  No evidence for a Higgs boson signal is observed, and we use
the null result to infer cross sections excluded at the 95\% C.L.  as
a function of the Higgs mass, and 95\% C.L. excluded regions of the MSSM
parameter space $\tan\beta$ versus $m_A$.


We thank A. Belyaev, M. Carena, J. Gunion, T. Han, S. Heinemeyer,
W. Kilgore, S. Mrenna, M. Spira, C. Wagner, G. Weiglein, and
S. Willenbrock for illuminating discussions on the theory of MSSM
Higgs production and decays.

We thank the Fermilab staff and the technical staffs of the
participating institutions for their vital contributions. This work
was supported by the U.S. Department of Energy and National Science
Foundation; the Italian Istituto Nazionale di Fisica Nucleare; the
Ministry of Education, Culture, Sports, Science and Technology of
Japan; the Natural Sciences and Engineering Research Council of
Canada; the National Science Council of the Republic of China; the
Swiss National Science Foundation; the A.P. Sloan Foundation; the
Bundesministerium f\"ur Bildung und Forschung, Germany; the Korean
Science and Engineering Foundation and the Korean Research Foundation;
the Science and Technology Facilities Council and the Royal Society,
UK; the Institut National de Physique Nucleaire et Physique des
Particules/CNRS; the Russian Foundation for Basic Research; the
Ministerio de Ciencia e Innovaci\'{o}n, and Programa
Consolider-Ingenio 2010, Spain; the Slovak R\&D Agency; and the
Academy of Finland.

\begin{table}[t!]
  \caption{Observed 95\% C.L. upper limits, and median expected limits under the null hypothesis on
           the Higgs boson production cross section times branching
           ratio $\sigma\cdot B$ versus $m_A$.}
  \begin{tabular}{ccc} \hline \hline
$m_A$ & median expected limit & observed limit \\  
GeV/$c^2$  &  (pb)     &    (pb)               \\ \hline
  90  &  28.115  &  28.978 \\
 100  &  19.884  &  23.465 \\
 110  &   9.382  &  11.063 \\
 120  &   5.447  &   5.288 \\
 130  &   3.374  &   2.770 \\
 140  &   2.340  &   1.812 \\
 150  &   1.751  &   1.392 \\
 160  &   1.400  &   1.198 \\
 170  &   1.124  &   1.051 \\
 180  &   0.933  &   0.880 \\
 190  &   0.782  &   0.808 \\
 200  &   0.707  &   0.709 \\
 230  &   0.470  &   0.505 \\
 250  &   0.379  &   0.451 \\ \hline \hline
  \end{tabular}
  \label{tab-limits}
\end{table}

\begin{figure}[t!]
  \begin{center}
     \includegraphics[width=\columnwidth]{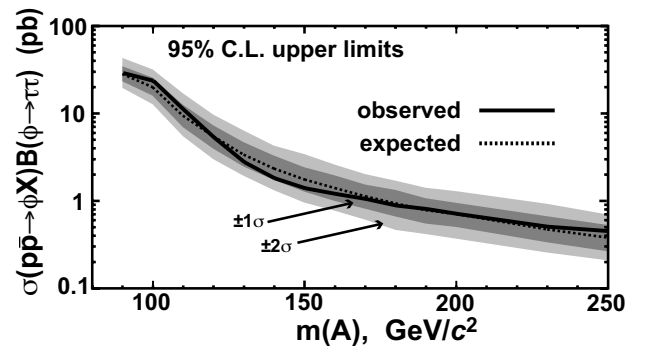}
  \end{center}
  \caption{Observed 95\% C.L. upper limits on the cross section 
           for $\phi=h/A/H$ production as a function of $m_A$.  The grey bands show
           the median expected limit under the null hypothesis, and indicate the 
           $\pm$1- and $\pm$2-standard-deviation ranges.}
  \label{fig-limit}
\end{figure}

\begin{figure}[t!]
  \begin{center}
     \includegraphics[width=\columnwidth]{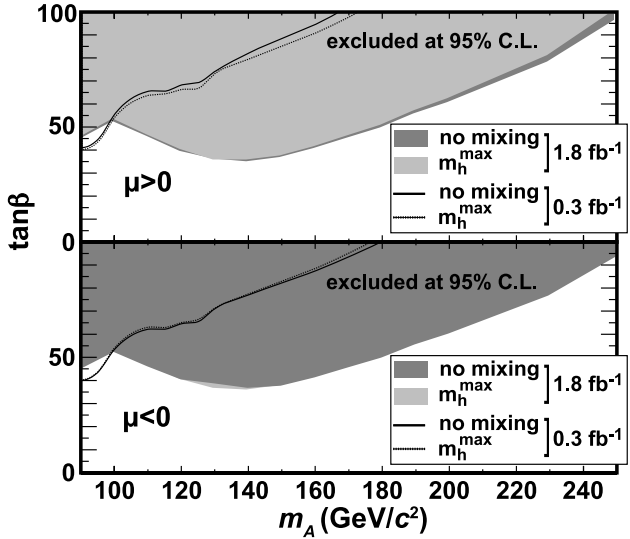}
  \end{center}
  \caption{Regions in the MSSM plane of $\tan\beta$ versus $m_A$ excluded at 95\% C.L.
           assuming heavy ($\sim$1 TeV/$c^2$) sfermions.  The top panel shows excluded regions
           for higgsino mass parameter $\mu>0$, and the bottom panel shows excluded regions
           for $\mu<0$.  Each panel shows the slightly different excluded regions for two scalar top mixing
           scenarios.  The solid and dashed curves show the previously published bounds.~\cite{Jan07}}
  \label{fig-tanb-ma}
\end{figure}



\begin{thebibliography}{99}
\bibitem{ref-eta} We use a coordinate system defined about the
    proton beam direction, which is taken as the $z$ axis; the $x$
    axis lies in the horizontal plane and the $y$ axis points upward.
    Then $\theta$ is the usual polar angle.  We define the
    pseudorapidity $\eta$ of a particle's three-momentum as $\eta
    \equiv -\ln(\tan\frac{\theta}{2})$.  The transverse energy and
    momentum are defined as $E_T = E \sin\theta$ and $p_T =
    p\sin\theta$ where $E$ is the energy measured by the calorimeter
    and $p$ is the momentum measured in the tracking system.  The
    missing transverse energy is defined as $\met = |-\sum_i E_T^i
    \vec{n}_i |$ where $\vec{n}_i$ is a unit vector perpendicular to
    the beam axis and pointing from the beam axis to the 
    $i^{\rm th}$ calorimeter tower.
\end{thebibliography}
\end{document}